\documentclass[aps,twocolumn]{revtex4}   
\usepackage{amsmath,amssymb}
\usepackage{bm}
\usepackage{epsfig}

\newcommand{\dg}{\dagger}
\newcommand{\ti}{\tilde}
\newcommand{\nl}{\nonumber \\}
\newcommand{\la}{\langle}
\newcommand{\ra}{\rangle}
\newcommand{\lla}{\la\!\la}
\newcommand{\rra}{\ra\!\ra}

\newcommand{\Sch}{Schr\"{o}dinger\ }

\newcommand{\Sec}[1]{Sec.\,\ref{#1}}

\newcommand{\be}{\begin{equation}}
\newcommand{\ee}{\end{equation}}
\newcommand{\bea}{\begin{eqnarray}}
\newcommand{\eea}{\end{eqnarray}}
\newcommand{\bsube}{\begin{subequations}}
\newcommand{\esube}{\end{subequations}}
\newcommand{\Eq}[1]{Eq.\,(\ref{#1})}
\newcommand{\Eqs}[1]{Eqs.\,(\ref{#1})}
\newcommand{\Fig}[1]{Fig.\,\ref{#1}}
\newcommand{\Figs}[1]{Figs.\,\ref{#1}}

\newcommand{\ind}{{\sf n}}
\newcommand{\indzero}{{\sf 0}}

\newcommand{\bfG}  {\mbox{\boldmath${\cal G}$}}
\newcommand{\bfV}  {\mbox{\boldmath${\cal D}$}}
\newcommand{\bfB}  {\mbox{\boldmath${\cal B}$}}
\newcommand{\bfmu} {\mbox{\boldmath${\mu}$}}
\newcommand{\bfLamb} {\bm\Lambda}

\newcommand{\B}{\mbox{\tiny B}}

\newcommand{\leftact}{\overset{\rightarrow}}
\newcommand{\rightact}{\overset{\leftarrow}}

\begin{document}

\title{
 Advancing hierarchical equations of motion for efficient
 evaluation of coherent two-dimensional spectroscopy
}

\author{Jian Xu,$^{1}$ Rui-Xue Xu,$^{2,\,\ast}$ Darius Abramavicius,$^{3,4}$
Houdao Zhang,$^{1}$
and YiJing Yan$^{1,2,4,\,}$} \email{rxxu@ustc.edu.cn; yyan@ust.hk}
\affiliation{$^1$Department of Chemistry,
 Hong Kong University of Science and Technology, Kowloon, Hong Kong SAR, China
}
\affiliation{$^2$Hefei National Laboratory for Physical Sciences at
the Microscale, University of Science and Technology of China, Hefei, Anhui 230026, China
}
\affiliation{$^3$Physics Faculty, Vilnius University, Lithuania
}
\affiliation{$^4$State Key Laboratory of Supramolecular Materials, Jilin University, China
}

\date{27 September 2011. Chinese Journal of Chemical Physics, in press}

\begin{abstract}
 To advance hierarchial equations of motion as a standard theory
for quantum dissipative dynamics,
we put forward a mixed Heisenberg--\Sch scheme with block-matrix implementation
on efficient evaluation of nonlinear optical response function.
The new approach is also integrated with optimized hierarchical theory and
numerical filtering algorithm.
Different configurations of coherent two-dimensional
spectroscopy of model excitonic dimer systems are
investigated, with focus on the effects of intermolecular transfer coupling
and bi-exciton interaction.

\end{abstract}

\maketitle

\section{Introduction}
\label{thintro}

 Coherent multi-dimensional spectroscopies provide powerful tools to investigate
various effects of molecular interactions and dynamic correlations
\cite{Muk95,Muk00691,Bri05625,Che09241,Abr092350,Muk09}.
Recent experiments show the evidence of long-lived
quantum coherence in photosynthesis systems
\cite{Eng07782,Lee071462,Cal0916291},
even at room temperatures
\cite{Pan1012766,Col10644}. 
Theoretical studies are also carried out towards the simulating and understanding
of the involved excitation energy transfer processes
\cite{Ish0917255,Muk042073,Ado062778,Abr088525,Che10024505,Che11194508}. 
In such systems, the pigment-protein interaction
has the same magnitude as the pigment-pigment transfer coupling,
and the time scale of environment memory is about
that of excitation energy transfer.
These characteristics require the non-Markovian
and non-perturbative quantum dissipation methods.
Among them, hierarchical equations of motion (HEOM) approach
\cite{Tan906676,Tan06082001,Yan04216,Sha045053,Xu05041103,Xu07031107}, 
an equivalence to the Feymann-Vernon influence functional path integral
\cite{Fey63118,Wei08,Kle09} 
but numerically more efficient alternative,
emerges as a standard method. %

There are many recent efforts, in
both theoretical and numerical aspects,
on advancing exact HEOM to be a power and versatile
tool for the study of various quantum dissipative systems.
HEOM formalism was originally proposed in 1989
by Tanimura and Kubo for semiclassical dissipation \cite{Tan89101}.
Formally exact HEOM formalism \cite{Tan906676,%
 Tan06082001,Yan04216,Sha045053,Xu05041103,Xu07031107} %
for Gaussian dissipation in general, including its second
quantization \cite{Jin08234703}, has now been well established.
The major obstacle of HEOM is its numerical tractability.
The number of equations involved in the theory are usually huge.
Brute-force implementation is greatly limited by both memory and processing capability.
One major numerical advancement is the on-the-fly filtering algorithm \cite{Shi09084105}.
It goes with a preselected error tolerance on the properly scaled HEOM.
The filtering algorithm dramatically reduces the effective number of equations,
and it by nature also automatically truncates the hierarchy level.
In the formulation front, the Pad\'{e} spectrum decomposition (PSD) scheme
\cite{Hu10101106,Hu11244106} has been proposed
for an optimized HEOM construction \cite{Ding11jcp}.
It dramatically reduces the number of equations,
in comparison with that of the conventional
Matsubara expansion based formalism at same accuracy.
A priori accuracy control criterion has also been proposed
\cite{Xu09214111,Tia10114112,Ding11jcp}
so that the optimized HEOM can be used confidently,
without costly convergency test.

In this work, we discuss two additional techniques
to further improve the efficiency of HEOM in evaluating such as
the third-order optical response functions and two-dimensional spectroscopy.
One is the mixed Heisenberg--\Sch scheme.
The conventional approach follows
the direct realization of third-order response function from the view of \Sch picture.
It propagates the reduced system density operator
in three nested time intervals, denoted as $t_1$, $t_2$, and $t_3$,
between three interrogations and the time of detection which is
through dynamic variable such as transition dipole.
In the new scheme, while retaining $t_1$- and $t_2$-propagations on state variables,
we unlash $t_3$ from the nested loop
via its propagation in Heisenberg picture on dynamic variables.
HEOM in \Sch picture and Heisenberg picture
are detailed in \Sec{theoaA} and \Sec{theoaB}, respectively.

Another advancement in this work is the block-HEOM dynamics,
for its efficient evaluation of  coherent two-dimensional spectroscopy, see \Sec{theob}.
In optical processes, the system Hamiltonian is considered
to be block diagonalized in electronic manifolds, in virtue of
Born-Oppenheimer principle.
We also assume that the relaxation between different manifolds is negligible.
The resulting HEOM dynamics will be in the block-matrix form,
even as they involve in the third-order optical response functions,
where the dynamics occur in population and/or coherence states in electronic space.
Combining the block-HEOM theory
with the mixed Heisenberg--\Sch scheme
greatly facilitates the evaluation of
third-order optical response functions.
Moreover, numerical filtering algorithm \cite{Shi09084105}
is now also extended to the present formalism.

 We exemplify our method with model excitonic dimer systems
in \Sec{num}.
The well-established ${\bf k}_{\rm I}$, ${\bf k}_{\rm II}$ and ${\bf k}_{\rm III}$
types of coherent two-dimensional spectroscopy \cite{Abr092350} are evaluated
with the present HEOM dynamics that integrates
all the state-of-the-art techniques.
Finally we summarize the paper in \Sec{thsum}.

\section{Hierarchical equations of motion}
\label{theoa}

\subsection{HEOM in \Sch picture}
\label{theoaA}

 HEOM couples the reduced system density operator
of primary interest, $\rho(t)\equiv{\rm tr}_{\B}\rho_{\rm total}(t)$,
to a set of auxiliary density operators (ADOs).
As an exact and nonperturbative theory,
HEOM accounts for the combined effects of
system-bath coupling strength, environment memory timescales,
and many-body interactions \cite{Xu07031107,Jin08234703}.
The explicit form of HEOM is
defined upon a certain statistical environment bath ``basis set''
that decomposes the interacting bath correlation functions
into distinct memory-frequency components.
Without loss of generality, we exemplify it
with the case of single-mode system-bath interaction,
$H'(t)=-QF_{\B}(t)$,
where $Q$ and $F_{\B}(t)$ are operators in
the reduced system and the stochastic bath subspaces, respectively.
In general, $H'(t)$ can be expressed in multiple-modes decomposition form.
The stochastic bath operator $F_{\B}(t)$ assumes a Gaussian process.
The influence of bath is therefore described via its correlation function
$C(t) \equiv \la F_{\B}(t)F_{\B}(0)\ra_{\B}$.
It is related to the bath spectral density $J(\omega)$
via the fluctuation-dissipation theorem \cite{Wei08,Yan05187}:
\be\label{fdt0}
 C(t) = \frac{1}{\pi} \int_{-\infty}^{\infty}\!\!d\omega
  \frac{e^{-i\omega t}J(\omega)}{1-e^{-\beta\omega}},
\ee
where $1/(1-e^{-\beta\omega})$ is Bose function
at inverse temperature $\beta=\hbar/(k_BT)$.
We set $\hbar=1$ hereafter.

 To construct a HEOM formalism,
one need to expand $C(t)$ in a finite exponential series,
on the basis of certain sum-over-poles scheme, together with
the Cauchy residue theorem of contour integration applied to \Eq{fdt0}.
In this work we adopt the Drude model,
\be\label{Jdrude}
  J(\omega)
 =\frac{2\lambda\gamma \omega}{\omega^2+\gamma^2}.
\ee
In this case, $[N/N]$ Pad\'{e} spectrum decomposition (PSD)
for Bose function is shown to be the best
\cite{Hu11244106,Hu10101106,Xu09214111,Tia10114112,Ding11jcp}.
It results in an exponential expansion of interacting
bath correlation function \cite{Ding11jcp},
\be\label{fdt}
 C(t) \approx \sum_{k=0}^{N} c_k e^{-\gamma_kt}+2\Delta_N\delta(t),
\ee
with
\be\label{DeltaN}
\Delta_N=\frac{\lambda\beta\gamma}{2(N+1)(2N+3)}.
\ee
The $k=0$ term with $\gamma_0\equiv\gamma$
is the Drude pole contribution, while other $N$ contributions with $k=1,\cdots,N$
are from the $[N/N]$ PSD Bose function poles $\{\gamma_{k\neq 0}\}$ that
are all positive and can be easily identified \cite{Hu11244106}.
The $\delta$-function term in \Eq{fdt}
re-sums the off-basis-set residue outside the finite sum-over-poles scheme.
This is the only approximation
involved not just in the bath correlation function,
but also in the resulting HEOM that goes therefore by
a convenient accuracy control criterion
prior to dynamics evaluation \cite{Xu09214111,Tia10114112,Ding11jcp}.

 The exponential expansion form of $C(t)$ dictates the
construction of HEOM. According to \Eq{fdt}
it reads explicitly as \cite{Xu05041103,Xu07031107,Shi09084105,Shi09164518}
\begin{align}\label{rndot}
\dot\rho_{\ind}=&
 -[i{\cal L}(t)+\gamma_{\ind}+\Delta_N{\cal Q}^2]\rho_{\ind}
\nl&
  -i\sum_{k=0}^{N}
    \sqrt{\frac{n_k}{|c_k|}}\,
     \bigl(  c_k Q \rho_{{\ind}_{k}^-}
         -c^\ast_k \rho_{{\ind}_{k}^-} Q \bigr)\nl
  & -i\sum_{k=0}^{N}
   \sqrt{(n_{k}+1)|c_k|}\,{\cal Q}\rho_{{\ind}_{k}^+}\,,
\end{align}
with ${\cal L}\rho_{\ind}\equiv [H,\rho_{\ind}]$ for the reduced system Liouvillian,
     ${\cal Q}\rho_{\ind}\equiv [Q,\rho_{\ind}]$, and
\be\label{gamn}
  \gamma_{\ind}=\sum_{k=0}^{N} n_k \gamma_k.
\ee
The ADO labeling index is now specified as $\ind\equiv\{n_0,n_1,\cdots,n_K\}$,
with $n_k\geq 0$.
It consists a set of non-negative indices,
following the exponential expansion of \Eq{fdt}.
The index ${\ind}_{k}^{\pm}$ in the last two terms of \Eq{rndot}
differs from $\ind$ only by changing
the specified $n_k$ to
$n_k\pm 1$.
Let $n_0+n_1+\cdots+n_K=n$ and call
the individual $\rho_{\ind}\equiv \rho_{n_0,n_1,\cdots,n_K}$ an $n^{\rm th}$-tier ADO.
It depends on its associated $(n\pm 1)^{\rm th}$-tier ADOs,
specified individually by the last two terms in \Eq{rndot}.
The reduced system density operator of primary interest is just the
zeroth-tier ADO, $\rho_{\indzero}(t)\equiv\rho(t)$.
The $\Delta_N$-term in \Eq{rndot} arises from the white-noise residue in
\Eq{fdt}. All ADOs in \Eq{rndot} are dimensionless and scaled properly
to support the efficient HEOM propagator via the on-the-fly
filtering algorithm that also automatically truncates the level of
hierarchy \cite{Shi09084105}. The nonperturbative nature of HEOM has
been discussed in detail, see Ref.~\onlinecite{Xu07031107} for
example.
HEOM (\ref{rndot}) is called to be in \Sch picture,
for its governing the system state variables, which
are the ADOs including the reduced system density operator of primary interest.

\subsection{HEOM in Heisenberg picture}
\label{theoaB}

 HEOM supports the evaluations of not only the expectation values
but also the correlation and response functions of any dynamics variables
of the reduced system.
Nonlinear optical response functions for two-dimensional spectroscopy
will be described in the next section.
They will be evaluated based on the mixed Heisenberg--\Sch scheme and
the block-HEOM dynamics.

Note that HEOM consists a set of linear coupled equations.
The HEOM space algebra is mathematically the same of linear space,
regardless its physics contents.
We recast HEOM as $\dot{\bm\rho}(t)=-\hat\bfLamb(t){\bm\rho}(t)$,
with the column vector $\bm\rho\equiv\{\rho_{\ind={\indzero}},
\rho_{\ind\neq\indzero}\}$ of ADOs defining a basic element
in HEOM space.
The dynamic generator $\hat\bfLamb(t)$
is determined by specific form of HEOM,
i.e.\ \Eq{rndot}.
The HEOM propagator $\hat\bfG(t,\tau)$ by which $\bm\rho(t)=\hat\bfG(t,\tau)\bm\rho(\tau)$
satisfies
$\partial\hat\bfG(t,\tau)/\partial t=-\hat\bfLamb(t)\hat\bfG(t,\tau)$.
In the absence of time-dependent external field,
${\cal L}(t)={\cal L}$ and therefore $\hat\bfLamb(t)=\bfLamb$
are time-independent, resulting in
$\hat{\bfG}(t,\tau)=\exp[-\bfLamb(t-\tau)]\equiv\bfG(t-\tau)$.

The expectation value of a system dynamical variable reads
\be\label{barA}
 \bar A= {\rm tr}(A\rho)\equiv \lla A|\rho\rra
 =\lla\bm A|\bm\rho\rra \equiv
 \sum_{\text{all $\ind$}} \lla A_{\ind}|\rho_{\ind}\rra.
\ee
The first two identities are the conventional
reduced Liouville space expressions.
The last two identities are the extensions to HEOM space,
in which, according to the third identity,
\be\label{A0}
{\bm A}\equiv\{A_{\ind=\indzero}=A,\,A_{\ind\neq\indzero}=0\}={\bm A}(0).
\ee
The last identity will be used later to specify
the initial values of HEOM in Heisenberg picture.

HEOM supports also the evaluation of correlation and response functions.
For example, we apply linear response theory to HEOM, resulting in
linear correlation function the expression of
\be\label{CorrAB}
  C_{AB}(t) =\lla{\bm A}|\bfG(t)\leftact{\bfB}|{\bm\rho}^{\rm eq}\rra ,
\ee
which is equivalent to
$C_{AB}(t) = \lla A|{\cal G}_M(t)|B\rho_M^{\rm eq}\rra=\la A(t)B(0)\ra$.
The latter is the conventional expression, defined in the
full system-plus-bath material space.
However, \Eq{CorrAB} is defined in HEOM space of the reduced system only.
The equilibrium
${\bm\rho}^{\rm eq}\equiv\{\rho^{\rm eq}_{\ind=\indzero},\rho^{\rm eq}_{\ind\neq\indzero}\}$
is the stationary solution to HEOM, i.e.\ $\bfLamb\bm\rho_{\rm eq}=0$,
together with the normalization condition of the primary reduced
system density operator Tr$\rho_{\indzero}=1$.
The resulting
$\{\rho^{\rm eq}_{\ind\neq\indzero}\neq 0\}$ in general
account for the initial system-bath correlations.
In \Eq{CorrAB}, $\leftact{\bfB}{\bm\rho}^{\rm eq}
\equiv\{B\rho^{\rm eq}_{\ind=\indzero},B\rho^{\rm eq}_{\ind\neq\indzero}\}$,
in comparison with the
full material space counterpart of $\leftact B\rho_M^{\rm eq}=B\rho_M^{\rm eq}$.
Denote also $\rightact{\bfB}{\bm\rho}\equiv\{\rho_{\ind=\indzero}B,\rho_{\ind\neq\indzero}B\}$
for later use; cf.\ \Eq{Rexpr}.

Apparently, the propagator $\bfG(t)$ can take action
on a system state variable, e.g.,
$\bfG(t)(\leftact{\bfB}{\bm\rho}^{\rm eq})$ in \Eq{CorrAB}.
This is the \Sch picture. It can also act from right to left on a dynamic variable, i.e.,
${\bm A}(t)\equiv {\bm A}\bfG(t)$.
This is the Heisenberg picture, the HEOM analogue of the
conventional $A(t)\equiv A{\cal G}_M(t)$ that satisfies
the Heisenberg equation, $\dot A=-iA{\cal L}_M=-i[A,H_{M}]$,
defined in the full system-plus-bath material space.

 HEOM in Heisenberg picture satisfies
$\dot{\bm A}(t)=-{\bm A}(t)\bfLamb$, with
${\bm A}(0)={\bm A}$ defined in \Eq{A0}.
Its explicit expressions can be obtained as follows.
Let us start with the identity,
\be\label{Heisen0}
  \lla {\bm A}|\bfLamb|{\bm\rho}\rra = \lla\tilde{\bm A}|{\bm\rho}\rra
  = \lla{\bm A}|\tilde{\bm\rho}\rra ,
\ee
where $\tilde{\bm A} \equiv {\bm A}\bfLamb$ and
    $\tilde{\bm\rho} \equiv \bfLamb{\bm\rho}$,
with $\bfLamb$ acting from right and left, respectively.
Note that  $\tilde{\bm A}=- \dot{\bm A}$ and
$\tilde{\bm\rho}=- \dot{\bm\rho}$\,.
From \Eq{rndot} we have
\begin{align}\label{dtbarASch}
 \lla{\bm A}|\tilde{\bm\rho}\rra=&
   \sum_{\text{all $\ind$}} \biggl\{ \lla A_{\ind}|
  i{\cal L}+\gamma_{\ind}+\Delta_N{\cal Q}^2|\rho_{\ind}\rra
\nl&  +i \sum_{k=0}^{N}
   \sqrt{\frac{n_k}{|c_k|}}\,
     \bigl(  c_k   \lla A_{\ind}|Q \rho_{{\ind}_{k}^-}\rra
         -c^\ast_k \lla A_{\ind}|\rho_{{\ind}_{k}^-} Q \rra\bigr)
\nl&  +i \sum_{k=0}^{N}
   \sqrt{(n_{k}+1)|c_k|}\,\lla A_{\ind}|{\cal Q}|\rho_{{\ind}_{k}^+}\rra \biggr\}.
\end{align}
In contact with the second quantity in \Eq{Heisen0},
we recast every individual term above with respect to
the same $\rho_{\ind}\equiv\rho_{n_0,n_1\cdots,n_N}$, by using  for example the identity,
$$
  \sum_{\text{all}\ \ind} \sqrt{\frac{n_k}{|c_k|}}\lla A_{\ind}|Q \rho_{{\ind}_{k}^-}\rra
 = \sum_{\text{all}\ \ind}\sqrt{\frac{n_k+1}{|c_k|}}\lla A_{\ind_{k}^+}Q|\rho_{\ind}\rra \,.
$$
We can therefore recast \Eq{dtbarASch} as
\begin{align*}
 \lla\tilde{\bm A}|{\bm\rho}\rra=&
   \sum_{\text{all $\ind$}} \biggl\{ \lla A_{\ind}|
  i{\cal L}+\gamma_{\ind}+\Delta_N{\cal Q}^2|\rho_{\ind}\rra
\nl& +i \sum_{k=0}^{N}\!
 \sqrt{\frac{n_k\!+\!1}{|c_k|}}\!
     \Bigl(c_k\lla A_{{\ind}_{k}^+} Q|\rho_{{\ind}}\rra\!
         -c^\ast_k \lla Q A_{{\ind}_{k}^+}|\rho_{{\ind}}\rra \Bigr)
\nl&
  +i \sum_{k=0}^{N}
   \sqrt{n_k|c_k|}\,\lla A_{{\ind}_{k}^-}|{\cal Q}|\rho_{{\ind}}\rra \biggr\}.
\end{align*}
Accordingly, HEOM in Heisenberg picture reads explicitly as \cite{Zhu115678Note}:
\begin{align}\label{Andot}
\dot A_{\ind}=&-A_{\ind}(i{\cal L}+ \gamma_{\ind} +\Delta_N {\cal Q}^2) \nl
  & -i\sum_{k=0}^{N}
    \sqrt{\frac{n_k+1}{|c_k|}}\,
     \bigl(  c_k A_{{\ind}_{k}^+} Q
         -c^\ast_k Q A_{{\ind}_{k}^+}  \bigr)\nl
  & -i\sum_{k=0}^{N}
   \sqrt{n_k|c_k|}\,A_{{\ind}_{k}^-}{\cal Q}\,.
\end{align}
Here, $O{\cal L}=[O,H]$ and
$O{\cal Q}=[O,Q]$,
following the identities
of  ${\cal L}O=[H,O]$ and
${\cal Q}O=[Q,O]$ defined earlier.
Apparently, \Eq{rndot} and \Eq{Andot} are equivalent
but just in different pictures.

\section{Efficient HEOM evaluation of two-dimensional spectroscopies}
\label{theob}

\subsection{Nonlinear optical response functions via block-HEOM dynamics}
\label{theobA}

 Now turn to the third-order optical response functions,
as probed by coherent two-dimensional spectroscopies, operated with short pulsed
fields in certain four-wave-mixing configurations \cite{Muk00691,Abr092350}.
 Following the similar algebra of the linear correlation/response
function, the third-order optical response function in HEOM space
is obtained to be
$$
 R^{(3)}(t_3,t_2,t_1)=\lla{\bfmu}_{{\bf k}_{\rm s}}|\bfG(t_3)\bfV_{{\bf k}_3}
      \bfG(t_2)\bfV_{{\bf k}_2}\bfG(t_1)\bfV_{{\bf k}_1}|\bm\rho^{\rm eq}\rra.
$$
It is just the HEOM space analogue of the conventional
full system-plus-bath material space expression \cite{Muk95}.
Apparently the mixed Heisenberg--\Sch scheme
as commented in \Sec{thintro} will greatly
facilitate the evaluation of third-order response function.

 As a kind of four-wave-mixing spectroscopy,
the wavevector ${\bf k}_s$ of the signal field
satisfies the phase-matching condition.
It is that ${\bf k}_s=\pm{\bf k}_3\pm{\bf k}_2\pm{\bf k}_1$,
in relation to the three incident pulsed fields
interacting with the system sequentially.
Three basic configurations of coherent two-dimensional
spectroscopy \cite{Muk00691,Abr092350}
will be classified in \Sec{theobB}.
Their efficient evaluation via
block-HEOM in mixed Heisenberg--\Sch
scheme will be detailed in \Sec{theobC}.

 In the following, we show that
the optical response function can be recast
in the block-HEOM dynamics form.
Consider the third-order optical processes
involving the initial ground $|g\ra$, the excited $|e\ra$, and doubly-excited
$|f\ra$ manifolds of electronic states.
Assume also that the relaxation between different
manifolds is negligible.
This implies that not only the system Hamiltonian
but also the dissipative mode $Q$ are
block diagonalized, in virtue of
Born-Oppenheimer principle.

\begin{figure}
\includegraphics[width=1.0\columnwidth]{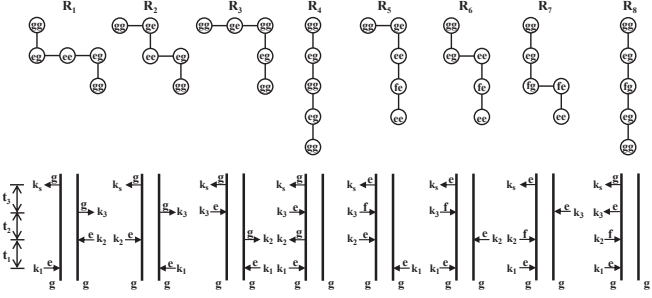}
\caption{Eight Liouville-space pathways (upper) and double-sided Feynman diagrams
(lower) for two-dimensional spectroscopy in the rotating wave approximation.
Each Liouville-space pathway starts from the
upper-left circle, while the double-sided Feynman diagram
starts from the bottom, following the
convention of Ref.\ \onlinecite{Muk95}.
 }
\label{fig1}
\end{figure}

On the other hand, the transition dipole operators involved in
the third-order optical response function are also in the block-matrix form,
although not diagonal. For example, the transition dipole
$\mu_-\equiv \hat\mu_{ge}|g\ra\la e| +\hat\mu_{ef}|e\ra\la f|$
amounts explicitly to
$$
 \mu_-=\begin{bmatrix}
  \hat 0_{gg} & \hat\mu_{ge} & \hat 0_{gf} \\
  \hat 0_{eg} & \hat 0_{ee} & \hat\mu_{ef} \\
  \hat 0_{fg} & \hat 0_{fe} & \hat 0_{ff}
 \end{bmatrix}.
$$
The matrix $\hat O_{uv}$ in the $(uv)$-block, with $u,v\in \{g,e,f\}$,
has the order of $N_{u}\times N_{v}$.
We have $N_g=1$, $N_e=M$ and $N_f=M(M-1)/2$,
for a molecular aggregate of size $M$, assuming
the simple exciton model.

 Expanding the third-order optical response function in the eight
Liouville-space pathways, as shown in \Fig{fig1},
and their complex conjugate counterparts leads to \cite{Yan885160,Muk95}
\be\label{R3in8}
  R^{(3)}(t_3,t_2,t_1)=i^3\sum_{\alpha=1}^8 [R_\alpha(t_3,t_2,t_1)-{\rm c.c.}].
\ee
These eight pathways contributions are expressed
in terms of the block-matrix dynamics
in HEOM space as
\begin{widetext}
\be\label{Rexpr}
\begin{split}
R_1(t_3,t_2,t_1)&=
   \lla \bfmu_{ge}|\bfG_{eg}(t_3)\rightact{\bfmu}_{eg}\bfG_{ee}(t_2) \rightact{\bfmu}_{ge}
   \bfG_{eg}(t_1)  \leftact{\bfmu}_{eg}|{\bm\rho}^{\rm eq}_{gg}\rra
  \, e^{-i\omega_{eg}(t_3+t_1)},
\\
R_2(t_3,t_2,t_1)&=
   \lla \bfmu_{ge}|\bfG_{eg}(t_3) \rightact{\bfmu}_{eg} \bfG_{ee}(t_2)  \leftact{\bfmu}_{eg}
   \bfG_{ge}(t_1) \rightact{\bfmu}_{ge}|{\bm\rho}^{\rm eq}_{gg}\rra
  \, e^{-i\omega_{eg}(t_3-t_1)},
\\
R_3(t_3,t_2,t_1)&=
   \lla \bfmu_{ge} |\bfG_{eg}(t_3) \leftact{\bfmu}_{eg} \bfG_{gg}(t_2) \rightact{\bfmu}_{eg}
   \bfG_{ge}(t_1) \rightact{\bfmu}_{ge}|{\bm\rho}^{\rm eq}_{gg}\rra
  \, e^{-i\omega_{eg}(t_3-t_1)},
\\
R_4(t_3,t_2,t_1)&=
   \lla \bfmu_{ge} |\bfG_{eg}(t_3) \leftact{\bfmu}_{eg} \bfG_{gg}(t_2)  \leftact{\bfmu}_{ge}
   \bfG_{eg}(t_1)  \leftact{\bfmu}_{eg}|{\bm\rho}^{\rm eq}_{gg}\rra
  \, e^{-i\omega_{eg}(t_3+t_1)},
\\
R_5(t_3,t_2,t_1)&= -
   \lla \bfmu_{ef} |\bfG_{fe}(t_3) \leftact{\bfmu}_{fe} \bfG_{ee}(t_2)  \leftact{\bfmu}_{eg}
   \bfG_{ge}(t_1) \rightact{\bfmu}_{ge}|{\bm\rho}^{\rm eq}_{gg}\rra
  \, e^{-i(\omega_{fe}t_3-\omega_{eg}t_1)},
\\
R_6(t_3,t_2,t_1)&= -
   \lla \bfmu_{ef} |\bfG_{fe}(t_3) \leftact{\bfmu}_{fe} \bfG_{ee}(t_2) \rightact{\bfmu}_{ge}
   \bfG_{eg}(t_1)  \leftact{\bfmu}_{eg}|{\bm\rho}^{\rm eq}_{gg}\rra
  \, e^{-i(\omega_{fe}t_3+\omega_{eg}t_1)},
\\
R_7(t_3,t_2,t_1)&= -
   \lla \bfmu_{ef}|\bfG_{fe}(t_3) \rightact{\bfmu}_{ge} \bfG_{fg}(t_2)  \leftact{\bfmu}_{fe}
   \bfG_{eg}(t_1)  \leftact{\bfmu}_{eg} |{\bm\rho}^{\rm eq}_{gg}\rra
  \, e^{-i(\omega_{fe}t_3+\omega_{fg}t_2+\omega_{eg}t_1)},
\\
R_8(t_3,t_2,t_1)&=
   \lla \bfmu_{ge} |\bfG_{eg}(t_3) \leftact{\bfmu}_{ef} \bfG_{fg}(t_2)  \leftact{\bfmu}_{fe}
   \bfG_{eg}(t_1)  \leftact{\bfmu}_{eg} |{\bm\rho}^{\rm eq}_{gg}\rra
  \, e^{-i(\omega_{eg}t_3+\omega_{fg}t_2+\omega_{eg}t_1)}.
\end{split}
\ee
\end{widetext}
The underlying optical processes will be discussed
in \Sec{theobB}.
The electronic phase factor in
each individual $R_{\alpha}$
can be formally absorbed into the involving
Green's functions, i.e., $\bfG_{uv}(t)e^{-i\omega_{uv}t}\rightarrow\bfG_{uv}(t)$,
with $\omega_{uv}\equiv\epsilon_u-\epsilon_v$ denoting the
chosen reference frequency for the optical transition  between
two specified electronic manifolds.
In the present notion, $\bfG_{uv}(t)$
involves only slow motion dynamics, as
the highly oscillatory optical frequency component is factorized out
for numerical advantage.
The corresponding block HEOM dynamics in both
\Sch and Heisenberg pictures will be detailed
in \Sec{theobC}.
There are other advantages for the present notion.
The overall electronic phase factor
can in fact be used to distinguish rephasing
versus non-rephasing optical processes \cite{Muk95,Yan885160,Yan91179}
and also to visualize the rotating wave approximation as seen below.

\subsection{Coherent two-dimensional spectroscopies}
\label{theobB}

 There are three basic configurations
of coherent two-dimensional spectroscopy, and their signals
are denoted as  $S_{{\bf k}_{\rm I}}$, $S_{{\bf k}_{\rm II}}$,
and $S_{{\bf k}_{\rm III}}$, respectively \cite{Abr092350}.
For simplicity we adopt the rotating-wave approximation and
the impulsive fields limit.

 The $S_{{\bf k}_{\rm I}}$ signal goes with
${\bf k}_s={\bf k}_3 + {\bf k}_2 -{\bf k}_1$,
the stimulated photon echo or rephasing configuration \cite{Yan91179},
while the $S_{{\bf k}_{\rm II}}$ signal goes
with ${\bf k}_s={\bf k}_3 - {\bf k}_2 +{\bf k}_1$
and is non-rephasing.
With the aid of the double-sided Feynman diagrams in \Fig{fig1},
these two signals are identified to be 
\begin{align}\label{S12}
  S_{{\bf k}_{\rm I/II}}(\omega_3,t_2,\omega_1)&= {\rm Re}
   \int_0^\infty\!\!dt_3\!\!\int_0^\infty\!\!dt_1
   e^{i(\omega_3t_3\mp\omega_1t_1)} \nl & \qquad \ \
  \times R_{{\bf k}_{\rm I/II}}(t_3,t_2,t_1),
\end{align}
and related respectively to
\be\label{Rk1k2}
\begin{split}
  &R_{{\bf k}_{\rm I}}=R_2+R_3+R_5 \qquad \text{\ \ \ (rephasing)}\, ,
\\
 &R_{{\bf k}_{\rm II}}=R_1+R_4+R_6 \qquad \text{(non-rephasing)}\, .
\end{split}
\ee
The rephasing versus non-rephasing
nature of individual $R_{\alpha}$ in \Eq{Rexpr} or \Eq{Rk1k2}
can be inferred easily from its overall
electronic phase factor \cite{Muk95,Yan885160,Yan91179}.
The signs associating with the frequencies $\omega_3$ and $\omega_1$
in \Eq{S12} are resulted from the incident fields in the
specified four wave mixing configuration  in the impulsive limit,
as implied in \Fig{fig1}.
These signs are just opposite to those in the electronic phase factor
of participating $R_{\alpha}$ contributions.
Thus, the participated pathway contributions
via the electronic rotating wave approximation
are also evident in \Eq{S12}.

 Experiments can also be performed in the configuration that
the pulsed ${\bf k}_2$-field is applied continuously not only
after but also before the ${\bf k}_1$-field.
The resulting signal amounts to
$S_{{\bf k}_{\rm I}+{\bf k}_{\rm II}} = S_{{\bf k}_{\rm I}} + S_{{\bf k}_{\rm II}}$.
It is in fact the pump-probe absorption configuration,
involving all the six pathways $R_{1}$ to $R_6$ contributions.
As inferred from \Eq{Rexpr},
these six pathways can be classified into the
{\it excited-state emission} ($R_1,R_2$),
{\it ground-state bleaching} ($R_3,R_4$),
and {\it excited-state absorption} ($R_5,R_6$) contributions.
In fact, the $t_1$ and $t_3$ represent the excitation
and detection time periods, and therefore,
the $\omega_1$ and $\omega_3$ in \Eq{S12}
are the excitation and detection frequencies,
respectively.
The $t_2$ denotes the waiting time, during which the
system is either in the excited or the ground
state manifold, with underlying dynamics
being governed by $\bfG_{ee}(t_2)$
or  $\bfG_{gg}(t_2)$, respectively; see \Eq{Rexpr}.

 The $S_{{\bf k}_{\rm III}}$ signal goes with
${\bf k}_s=-{\bf k}_3 + {\bf k}_2 +{\bf k}_1$,
the double-excitation configuration, and is related to
\be\label{Rk3}
  R_{{\bf k}_{\rm III}}=R_7+R_8,
\ee
as inferred from the double-sided Feynman diagrams
in \Fig{fig1}. The $R_7$ and $R_8$ are the
{\it double-excitation absorption} pathways,
involving the
$|f\ra \leftarrow |e\ra \leftarrow |g\ra$ processes,
while the bra state remains in $\la g|$.
During the time $t_2$ period,
the ${\bf k}_{\rm III}$ configuration
explores therefore double quantum coherence
dynamics governed by $\bfG_{fg}(t_2)$,
in contrast to the ${\bf k}_{\rm I/II}$ scheme
involving electronic state population dynamics.
The detection ${\bf k}_3$ field
involves single-excitation absorption of
$\la e| \leftarrow\la g|$ in $R_7$
and single-excitation emission
$|e\ra \leftarrow |f\ra$ in $R_8$, as evident in
the corresponding double-sided Feynman diagrams in \Fig{fig1}.
The above analysis justifies the
fact that the ${\bf k}_{\rm III}$-signal is designed to
probe the correlation between single
and double excitations \cite{Muk02327,Muk07221105,Abr088525}.
The two-dimensional half-Fourier transforms
are therefore performed with $t_2$ to resolve
the double-excitation frequency
and with either $t_3$ or  $t_1$ to
resolve the specified single-excitation frequency.
In this work, we choose
\begin{align}\label{S3}
 S_{\bf k_{\rm III}}(\omega_3, \omega_2, t_1) = & \mbox{Re}
   \int_0^\infty\!\! dt_3\!\! \int_0^\infty\!\! dt_2
 e^{i(\omega_3 t_3 + \omega_2 t_2)}
\nl & \qquad \times
 R_{\bf k_{\rm III}} (t_3,t_2,t_1) \,.
\end{align}
In the independent exciton limit,
the ${\bf k}_{\rm III}$-signal vanishes.
This can be seen from the involving
$R_7$ and $R_8$ contributions [cf.\ \Eq{Rexpr}].
Besides the signs, these two contributions
differ by their single coherence dynamics in the $t_3$ period,
which is $\bfG_{fe}(t_3)$ in $R_7$
but $\bfG_{eg}(t_3)$ in $R_8$.
These two contributions would cancel each other
in the absence of both inter-exciton transfer coupling
and double-exciton correlation.
Thus the ${\bf k}_{\rm III}$ technique serves as a
sensitive probe for interactions between excitons.

\subsection{Implementation with block-HEOM in mixed Heisenberg--\Sch picture}
\label{theobC}

To implement the third order optical response functions in \Eq{Rexpr},
we start with the thermal equilibrium ${\bm\rho}^{\rm eq}_{gg}$
in the ground-state $|g\ra$-manifold.
As described earlier, it is determined
by the steady state solution to HEOM,
involving now only the $(gg)$-block part.
For the simple exciton model, the $|g\ra$-manifold contains only one level,
and the ADOs in $(gg)$-block are all $1\times 1$ matrices,
resulting in
${\bm\rho}^{\rm eq}_{gg}=\{\rho^{gg,{\rm eq}}_{\ind=\indzero}=1,
\rho^{gg,{\rm eq}}_{\ind\neq\indzero}=0\}$.

 Block-matrix multiplications are then followed:
\be\label{t10}
\leftact{\bfmu}_{eg}{\bm\rho}^{\rm eq}_{gg}
 =\{\hat\mu_{eg}\rho^{gg,{\rm eq}}_{\ind=\indzero},
 \hat\mu_{eg}\rho^{gg,{\rm eq}}_{\ind\neq\indzero}\}\equiv \tilde{\bm\rho}_{eg}(0).
\ee
Denote also $\tilde{\bm\rho}_{ge}(0) \equiv \rightact{\bfmu}_{ge}{\bm\rho}^{\rm eq}_{gg}
=\{\rho^{gg,{\rm eq}}_{\ind=\indzero}\hat \mu_{ge},
\rho^{gg,{\rm eq}}_{\ind\neq\indzero}\hat \mu_{ge}\}$.
They are the initial states
for the block-HEOM $\bfG_{uv}(t_1)$ propagations in
\Eq{Rexpr}.
Each ADO in $\tilde{\bm\rho}_{uv}$
is an $N_u\times N_v$ matrix, as inferred
in the ($uv$)-indexes, and also Hermite conjugate
with its counterpart
   in $\tilde{\bm\rho}_{vu}$;
i.e., $\tilde{\bm\rho}_{vu}
 =\{ \tilde{\rho}^{vu}_{\ind}     \}
 =\{(\tilde{\rho}^{uv}_{\ind})^\dg\}
 \equiv\tilde{\bm\rho}_{uv}^\dg$.

The $t_1$- and $t_2$-propagations in \Eq{Rexpr} are implemented in a nested
manner in \Sch picture.
This picture is defined via the action-from-left of $\bfG_{uv}(t)$
on state variables; e.g.,
$\tilde{\bm\rho}_{uv}(t)=\bfG_{uv}(t)\tilde{\bm\rho}_{uv}(0)$.
The block-HEOM in \Sch picture can be reduced from \Eq{rndot} as
\begin{align}\label{bHEOM-Sch}
 \dot{\ti\rho}^{uv}_{\ind}=& -i({\cal L}_{uv}+\gamma_{\ind}
   +\Delta_N{\cal Q}^2_{uv}){\ti\rho}^{uv}_{\ind}
 \nl &
   -i\sum_{k=0}^{N}
    \sqrt{\frac{n_k}{|c_k|}}\,
     \bigl(c_k Q_{uu} {\ti\rho}^{uv}_{{\ind}_{k}^-}
       -c^\ast_k{\ti\rho}^{uv}_{{\ind}_{k}^-} Q_{vv}\bigr)
 \nl &
  -i\sum_{k=0}^{N}
   \sqrt{(n_{k}+1)|c_k|}\,{\cal Q}_{uv}{\ti\rho}^{uv}_{{\ind}_{k}^+}\,.
\end{align}
Here,
${\cal L}_{uv}{\hat O}_{uv}\equiv H_{uu}{\hat O}_{uv}-{\hat O}_{uv}H_{vv}$
and ${\cal Q}_{uv}{\hat O}_{uv}\equiv
Q_{uu}{\hat O}_{uv}-{\hat O}_{uv}Q_{vv}$.
Equivalently,
${\hat O}_{vu}{\cal L}_{uv}={\hat O}_{vu}H_{uu}-H_{vv}{\hat O}_{vu}$
and
${\hat O}_{vu}{\cal Q}_{uv}={\hat O}_{vu}Q_{uu}-Q_{vv}{\hat O}_{vu}$,
which will be used in the following Heisenberg picture.

The $t_3$-propagation in \Eq{Rexpr} is implemented in Heisenberg picture,
in parallel with the $t_1$- and $t_2$-propagations.
The Heisenberg picture is defined via the action-from-right of $\bfG_{uv}(t)$
on dynamic variables,
i.e.,
${\bm A}_{vu}(t)={\bm A}_{vu}\bfG_{uv}(t)$.
The initial condition is
${\bm A}_{vu}(0)={\bm A}_{vu}=\{{\bm A}^{vu}_{\ind=\indzero}=A_{vu},
{\bm A}^{vu}_{\ind\neq\indzero}=0\}$, following \Eq{A0}.
In consistent with \Eq{bHEOM-Sch}
or \Eq{Andot}, the block-HEOM in Heisenberg picture reads
\begin{align}\label{bHEOM-Hei}
 \dot{A}^{vu}_{\ind}=& -i{A}^{vu}_{\ind}
  ({\cal L}_{uv}+\gamma_{\ind}
   +\Delta_N{\cal Q}^2_{uv})
 \nl &
   -i\sum_{k=0}^{N}
    \sqrt{\frac{n_k+1}{|c_k|}}\,
     \bigl(c_k {A}^{vu}_{{\ind}_{k}^+} Q_{uu}
       -c^\ast_k Q_{vv} {A}^{vu}_{{\ind}_{k}^+}\bigr)
 \nl &
  -i\sum_{k=0}^{N}
   \sqrt{n_k|c_k|}\,{A}^{vu}_{{\ind}_{k}^-}{\cal Q}_{uv}\,.
\end{align}

To evaluate the third-order optical response functions in \Eq{Rexpr}
with block-HEOM in mixed Heisenberg--\Sch scheme, we introduce
\be\label{t2t1}
 \ti{\bm\rho}_{uv}(t_2;t_1)\equiv \bfG_{uv}(t_2)\ti{\bm\rho}_{uv}(0;t_1),
\ee
for three types of initial $t_2$ conditions:
\be\label{t20}
\begin{split}
 \ti{\bm\rho}_{ee}(0;t_1)&=\rightact{\bfmu}_{ge}\ti{\bm\rho}_{eg}(t_1),
\\
 \ti{\bm\rho}_{gg}(0;t_1)&=\leftact{\bfmu}_{ge}\ti{\bm\rho}_{eg}(t_1),
\\
 \ti{\bm\rho}_{fg}(0;t_1)&=\leftact{\bfmu}_{fe}\ti{\bm\rho}_{eg}(t_1).
\end{split}
\ee
We can recast  \Eq{Rexpr} as (up to the phase factors)
\be\label{finalR}
\begin{split}
 R_1(t_3,t_2,t_1)&=\lla\bfmu_{ge}(t_3)|\rightact{\bfmu}_{eg}
   \ti{\bm\rho}_{ee}(t_2;t_1)\rra,
\\
 R_2(t_3,t_2,t_1)&=\lla\bfmu_{ge}(t_3)|\rightact{\bfmu}_{eg}
   \ti{\bm\rho}^{\dg}_{ee}(t_2;t_1)\rra,
\\
 R_3(t_3,t_2,t_1)&=\lla\bfmu_{ge}(t_3)|\leftact{\bfmu}_{eg}
   \ti{\bm\rho}^{\dg}_{gg}(t_2;t_1)\rra,
\\
 R_4(t_3,t_2,t_1)&=\lla\bfmu_{ge}(t_3)|\leftact{\bfmu}_{eg}
   \ti{\bm\rho}_{gg}(t_2;t_1)\rra,
\\
 R_5(t_3,t_2,t_1)&=-\lla\bfmu_{ef}(t_3)|\leftact{\bfmu}_{fe}
   \ti{\bm\rho}^{\dg}_{ee}(t_2;t_1)\rra,
\\
 R_6(t_3,t_2,t_1)&=-\lla\bfmu_{ef}(t_3)|\leftact{\bfmu}_{fe}
   \ti{\bm\rho}_{ee}(t_2;t_1)\rra,
\\
 R_7(t_3,t_2,t_1)&=-\lla\bfmu_{ef}(t_3)|\rightact{\bfmu}_{ge}
   \ti{\bm\rho}_{fg}(t_2;t_1)\rra,
\\
 R_8(t_3,t_2,t_1)&=\lla\bfmu_{ge}(t_3)|\leftact{\bfmu}_{ef}
   \ti{\bm\rho}_{fg}(t_2;t_1)\rra.
\end{split}
\ee
These are the final expressions for the mixed
Heisenberg--\Sch scheme block-HEOM
evaluation of third-order optical response functions
and coherent two-dimensional spectrums via such as \Eqs{S12} and (\ref{S3}).

 We have also successfully extended the on-the-fly filtering
algorithm \cite{Shi09084105}
to the block-HEOM dynamics in \Eq{finalR}.
There involve
the nested ($t_2;t_1$)-propagation
in \Sch picture [\Eq{bHEOM-Sch}]
and the separated $t_3$-propagation in Heisenberg picture
[\Eq{bHEOM-Hei}].
Setting the filtering error tolerance at $2\times10^{-5}$ is
found to be sufficient for HEOM dynamics in both \Sch and Heisenberg
pictures, as tested extensively on
various systems, with numerical accuracy by eyes.

\section{Numerical demonstrations}
\label{num}

  We exemplify the efficient evaluation of
coherent two-dimensional spectrums with a model excitonic dimer.
Its Hamiltonian reads
$$
 H =\epsilon_{1}\hat{b}^{\dg}_1\hat{b}_1
  +\epsilon_{2}\hat{b}^{\dg}_2\hat{b}_2
  +V(\hat{b}^{\dg}_1 \hat{b}_2+\hat{b}^{\dg}_2 \hat{b}_1)
   +U\hat{b}^{\dg}_1\hat{b}_1\hat{b}^{\dg}_2\hat{b}_2,
$$
where $b^{\dg}_m$  ($b_m$) denotes
the exciton creation (annihilation) operator
on the specified molecular site.
The model system consists of a total four levels:
$|g\ra=|00\ra$ in the ground-state manifold,
$|e\ra=|10\ra$ and $|01\ra$ in the single-exciton manifold,
and $|f\ra=|11\ra$ in the double-exciton manifold.
The Rabi frequency within the single-exciton
manifold is $\sqrt{(\epsilon_{1}-\epsilon_{2})^2+4V^2}$.
The electronic transition dipoles
$\mu_{eg}=(\mu_{1}|10\ra+\mu_{2}|01\ra)\la 00|$ and
$\mu_{fe}=|11\ra(\la 10|\mu_{2}+\la 01|\mu_{1})$.
 Co-linear $(xxxx)$ field polarization configuration is adopted,
so that the effect of dipole directions on
spectroscopic signals can be neglected.

Each on-site transition energy experiences fluctuations,
brought in through $Q_m=\hat b^{\dg}_m\hat b_m$
influence of bath in Drude model.
We neglect the cross correlation between
different on-site fluctuations,
and also the static disorders that
are irrelevant to the methodology of this work.
In the following, we set $\lambda= 60\,{\rm cm}^{-1}$ and $\gamma^{-1}= 100\,{\rm fs}$
for each individual on-site Drude dissipation [\Eq{Jdrude}]
and 77\,K for temperature. In all cases, the $[1/1]$-PSD scheme is sufficient
according to the established accuracy control criterion
\cite{Xu09214111,Tia10114112,Ding11jcp}.

\begin{figure}
\includegraphics[width=1.0\columnwidth]{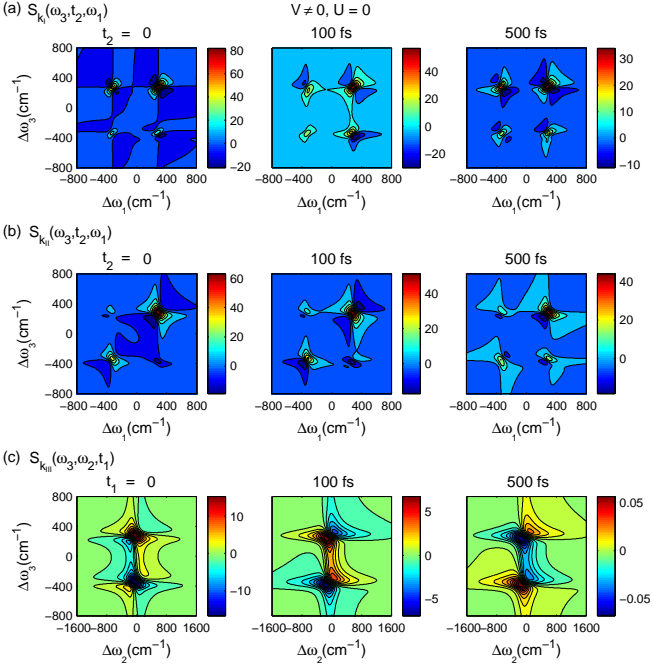}
\caption{Coherent two-dimensional spectra for the dimer system:
$\epsilon_1=\epsilon_2=\epsilon$, $V=-300\,{\rm cm}^{-1}$, and $U=0$, with
$\mu_1/\mu_2=-5$, at temperature 77\,K. Drude dissipation parameters
for each on-site excitation energy fluctuation
are $\lambda=60\,{\rm cm}^{-1}$ and $\gamma^{-1}=100$\,fs.
Frequencies (when applicable) are reported in terms of $\Delta\omega_1=\omega_1-\epsilon$,
$\Delta\omega_3=\omega_3-\epsilon$ and $\Delta\omega_2=\omega_2-2\epsilon$,
as the electronic reference transition frequencies are chosen to
be $\omega_{eg}=\omega_{fe}=\epsilon$
and $\omega_{fg}=2\epsilon$.
}
\label{fig2}
\end{figure}

\begin{figure}
\includegraphics[width=1.0\columnwidth]{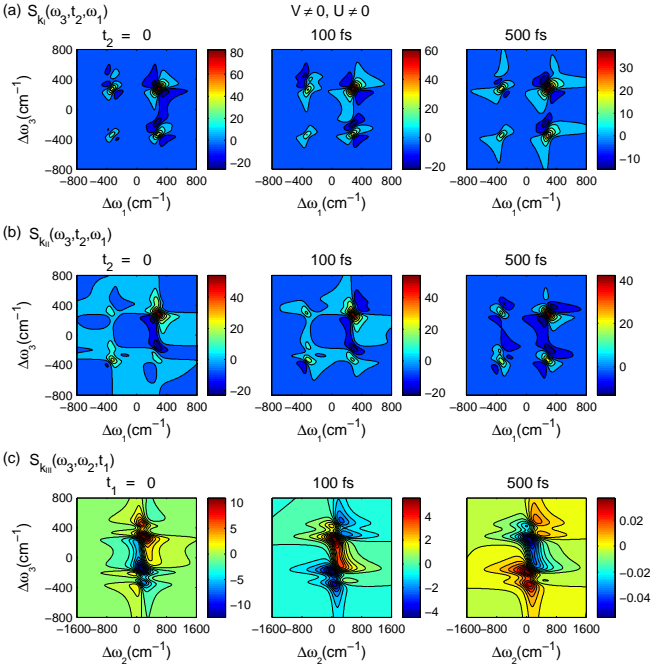}
\caption{Same as \Fig{fig2} but with a finite bi-exciton interaction,
$U=200\,{\rm cm}^{-1}$.}
\label{fig3}
\end{figure}

\begin{figure}
\includegraphics[width=1.0\columnwidth]{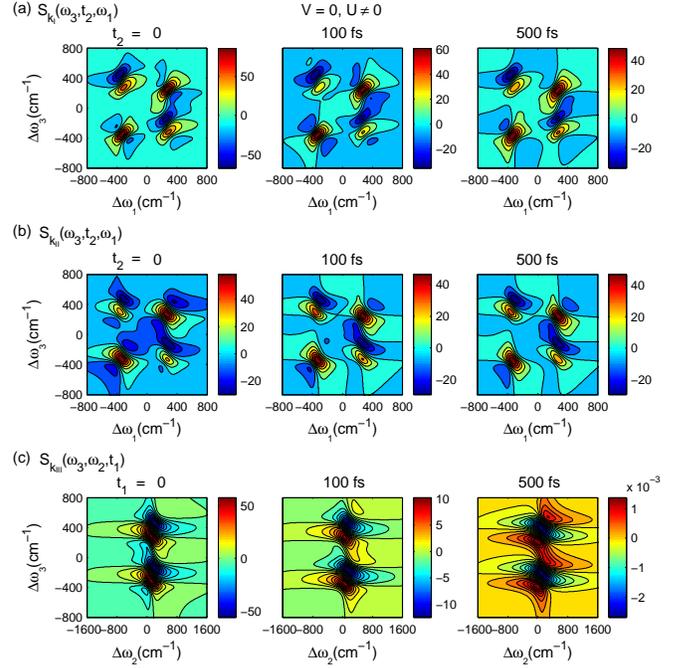}
\caption{Same as \Fig{fig3} but for
$\epsilon_1=\epsilon-300\,{\rm cm}^{-1}$, $\epsilon_2=\epsilon+300\,{\rm cm}^{-1}$,
and $V=0$, with $\mu_1/\mu_2=1$.}
\label{fig4}
\end{figure}

Figures \ref{fig2}--\ref{fig4} exemplify three representative cases of
$(V\neq 0,U=0)$, $(V\neq 0,U\neq 0)$, and $(V=0,U\neq 0)$, respectively,
but sharing a common value of Rabi frequency $\sqrt{(\epsilon_1-\epsilon_2)^2+4V^2}$.
Thus the peaks and valleys, which reflect
the transition frequencies between nonlocal eigenstates,
are distributed at similar positions in these three figures.
Highlighted are therefore
the effects of excitonic
transfer coupling $V$ and bi-exciton interaction $U$ on
$S_{{\bf k}_{\rm I}}$, $S_{{\bf k}_{\rm II}}$, and $S_{{\bf k}_{\rm III}}$,
as depicted in the panels (a), (b) and (c) of
each individual figure.
Intensities are reported in their relative values,
with a common factor over all frames in these figures.

 General speaking, both $V$ and $U$ affect correlations
between different monomers, manifested
via the cross peaks in  $S_{{\bf k}_{\rm I}}$
and $S_{{\bf k}_{\rm II}}$. On the other hand,
 $S_{{\bf k}_{\rm III}}$ that vanishes when $U=V=0$ is specialized
for the correlation between single and double excitation
coherence. It provides a different way to visualize the
effects of finite $U$ and/or $V$, as evident from \Figs{fig2}--\ref{fig4}.
The presence of $U$ manifests mainly on the
separation of peaks and valleys (cf.\ \Fig{fig4}
for example) and it is particularly prominent
in the  $S_{{\bf k}_{\rm III}}$ spectrum.

Other observations in rephasing $S_{{\bf k}_{\rm I}}$ and non-rephasing
$S_{{\bf k}_{\rm II}}$ are briefed as follows:
({\it i})
The peaks arise from excited state emission ($R_2$ and $R_1$) and ground state bleaching ($R_3$ and $R_4$),
 while the valleys are from
excited state absorption ($R_5$ and $R_6$);
({\it ii})
 The peaks/valleys are along the diagonal
direction in  $S_{{\bf k}_{\rm I}}$
and anti-diagonal  in  $S_{{\bf k}_{\rm II}}$,
in line with \Eq{S12}.
Inhomogeneity that is not included in calculations
affects mainly the diagonal direction.
Thus, it elongates the diagonal peaks in  $S_{{\bf k}_{\rm I}}$,
while  smears those in $S_{{\bf k}_{\rm II}}$;
({\it iii})
Bi-exciton interaction $U$
shifts the excited state absorption,
with respect to the excited state emission and ground state bleaching.
These two components differ in signs, being of valley versus peak.
Thus, cancelation would occur at least partially when $U=0$;
({\it iv})
Excitation energy transfer is observed
as the evolution of peaks/valleys intensities.
This process is mainly responsible by $V\neq 0$;
({\it v})
 The correlation effects arising from $U$ are separated out in \Fig{fig4}.
The negative peaks in \Fig{fig4} (a) and (b)
would cancel completely with the positive ones when $U=0$.
The larger the $U$ is, the bluer shift of the negative peaks
from their positive counterparts.
Note that the dimer system studied in \Fig{fig4} is nondegenerate,
rather than the degenerate ones in \Figs{fig2} and \ref{fig3},
for the appearance of correlation and coherence
between two distinct monomers.

\section{Concluding remarks}
\label{thsum}
 In summary, we propose a mixed Heisenberg--\Sch scheme and block-HEOM theory,
and demonstrate it with efficient evaluation of third-order optical response function
and coherent two-dimensional spectroscopy.
The new development has also been integrated with
the efficient numerical filtering algorithm \cite{Shi09084105}
and the optimized hierarchical theory \cite{Xu09214111,Tia10114112,Ding11jcp}. This is
the state-of-the-art HEOM approach.
For example, the calculations of all frames in \Fig{fig2} take
only about three minutes of CPU time
on a single processor of Intel(R) Xeon(R) E5472 (3GHz).
The development made in work will greatly facilitate the use of
HEOM, an exact and nonperbative quantum dissipation theory,
to the study of realistic systems.

\acknowledgments
 Support from the NNSF of China (21033008 \& 21073169),
the National Basic Research Program of China
(2010CB923300 \& 2011CB921400), and
the Hong Kong RGC (604709) and UGC (AoE/P-04/08-2)
is gratefully acknowledged.


\begin{thebibliography}{10}

\bibitem{Muk95}
S.~Mukamel,
\newblock {\em The Principles of Nonlinear Optical Spectroscopy},
\newblock Oxford University Press, New York, 1995.

\bibitem{Muk00691}
S.~Mukamel,
\newblock Annu. Rev. Phys. Chem. {\bf 51}, 691 (2000).

\bibitem{Bri05625}
T.~Brixner, J.~Stenger, H.~M. Vaswani,
   M.~Cho, R.~E. Blankenship, and G.~R. Fleming,
\newblock Nature {\bf 434}, 625 (2005).

\bibitem{Che09241}
Y.-C. Cheng and G.~R. Fleming,
\newblock Annu. Rev. Phys. Chem. {\bf 60}, 241 (2009).

\bibitem{Abr092350}
D.~Abramavicius, B.~Palmieri, D.~V. Voronine, F.~{$\check{\rm S}$}anda, and
  S.~Mukamel,
\newblock Chem. Rev. {\bf 109}, 2350 (2009).

\bibitem{Muk09}
 {\em Coherent Multidimensional Optical Spectroscopy}, edited by S.~Mukamel,
  Y.~Tanimura, and P.~Hamm, pages 1207--1469, Special Issue of Acc. Chem. Res.,
  Volume 42, Issue 9, 2009.

\bibitem{Eng07782}
G.~S. Engel, T.~R. Calhoun, E.~L. Read,
      T.~K. Ahn, T.~Man\v{c}al, Y.-C. Cheng,
      R.~E. Blankenship, and G.~R. Fleming,
\newblock Nature {\bf 446}, 782 (2007).

\bibitem{Lee071462}
H.~Lee, Y.-C. Cheng, and G.~R. Fleming,
\newblock Science {\bf 316}, 1462 (2007).

\bibitem{Cal0916291}
T.~R. Calhoun, N.~S. Ginsberg, G.~S. Schlau-Cohen, Y.-C. Cheng,
  M.~Ballottari, R.~Bassi, and G.~R. Fleming,
\newblock J. Phys. Chem. B {\bf 113}, 16291 (2009).

\bibitem{Pan1012766}
G.~Panitchayangkoon, D.~Hayes,
   K.~A. Fransted, J.~R. Caram, E.~Harel,
   J.~Z. Wen, R.~E. Blankenship, and G.~S. Engel,
\newblock Proc. Natl. Acad. Sci. USA {\bf 107}, 12766 (2010).

\bibitem{Col10644}
E.~Collini, C.~Y. Wong, K.~E. Wilk, P.~M.~G. Curmi, P.~Brumer, and G.~D. Scholes
\newblock Nature {\bf 463}, 644 (2010).

\bibitem{Ish0917255}
A.~Ishizaki and G.~R. Fleming,
\newblock Proc.\ Natl.\ Acad.\ Sci.\ USA {\bf 106}, 17255 (2009).

\bibitem{Muk042073}
S.~Mukamel and D.~Abramavicius,
\newblock Chem. Rev. {\bf 108}, 2073 (2004).

\bibitem{Ado062778}
J.~Adolphs and T.~Renger,
\newblock Biophys. J. {\bf 91}, 2778  (2006).

\bibitem{Abr088525}
D.~Abramavicius, D.~V. Voronine, and S.~Mukamel,
\newblock Proc. Natl. Acad. Sci. USA {\bf 105}, 8525 (2008).

\bibitem{Che10024505}
L.~P. Chen, R.~H. Zheng, Q.~Shi, and Y.~J. Yan,
\newblock J. Chem. Phys. {\bf 132}, 024505 (2010).

\bibitem{Che11194508}
L.~P. Chen, R.~H. Zheng, Y.~Y. Jing, and Q.~Shi,
\newblock J. Chem. Phys. {\bf 134}, 194508 (2011).

\bibitem{Tan906676}
Y.~Tanimura,
\newblock Phys. Rev. A {\bf 41}, 6676 (1990).

\bibitem{Tan06082001}
Y.~Tanimura,
\newblock J. Phys. Soc. Jpn. {\bf 75}, 082001 (2006).

\bibitem{Yan04216}
Y.~A. Yan, F.~Yang, Y.~Liu, and J.~S. Shao,
\newblock Chem. Phys. Lett. {\bf 395}, 216 (2004).

\bibitem{Sha045053}
J.~S. Shao,
\newblock J. Chem. Phys. {\bf 120}, 5053 (2004).

\bibitem{Xu05041103}
R.~X. Xu, P.~Cui, X.~Q. Li, Y.~Mo, and Y.~J. Yan,
\newblock J. Chem. Phys. {\bf 122}, 041103 (2005).

\bibitem{Xu07031107}
R.~X. Xu and Y.~J. Yan,
\newblock Phys. Rev. E {\bf 75}, 031107 (2007).

\bibitem{Fey63118}
R.~P. Feynman and F.~L. \mbox{Vernon, Jr.},
\newblock Ann. Phys. {\bf 24}, 118 (1963).

\bibitem{Wei08}
U.~Weiss,
\newblock {\em Quantum Dissipative Systems},
\newblock World Scientific, Singapore, 2008,
\newblock 3rd ed. Series in Modern Condensed Matter Physics, Vol.\ 13.

\bibitem{Kle09}
H.~Kleinert,
\newblock {\em Path Integrals in Quantum Mechanics, Statistics, Polymer
  Physics, and Financial Markets},
\newblock World Scientific, Singapore, 2009,
\newblock 5th ed.

\bibitem{Tan89101}
Y.~Tanimura and R.~Kubo,
\newblock J. Phys. Soc. Jpn. {\bf 58}, 101 (1989).

\bibitem{Jin08234703}
J.~S. Jin, X.~Zheng, and Y.~J. Yan,
\newblock J. Chem. Phys. {\bf 128}, 234703 (2008).

\bibitem{Shi09084105}
Q.~Shi, L.~P. Chen, G.~J. Nan, R.~X. Xu, and Y.~J. Yan,
\newblock J. Chem. Phys. {\bf 130}, 084105 (2009).

\bibitem{Hu10101106}
J.~Hu, R.~X. Xu, and Y.~J. Yan,
\newblock J. Chem. Phys. {\bf 133}, 101106 (2010).

\bibitem{Hu11244106}
J.~Hu, M.~Luo, F.~Jiang, R.~X. Xu, and Y.~J. Yan,
\newblock J. Chem. Phys. {\bf 134}, 244106 (2011).

\bibitem{Ding11jcp}
J.~J. Ding, J.~Xu, J.~Hu, R.~X. Xu, and Y.~J. Yan,
\newblock J. Chem. Phys.  (2011),
\newblock accepted.

\bibitem{Xu09214111}
R.~X. Xu, B.~L. Tian, J.~Xu, Q.~Shi, and Y.~J. Yan,
\newblock J. Chem. Phys. {\bf 131}, 214111 (2009).

\bibitem{Tia10114112}
B.~L. Tian, J.~J. Ding, R.~X. Xu, and Y.~J. Yan,
\newblock J. Chem. Phys. {\bf 133}, 114112 (2010).

\bibitem{Yan05187}
Y.~J. Yan and R.~X. Xu,
\newblock Annu. Rev. Phys. Chem. {\bf 56}, 187 (2005).

\bibitem{Shi09164518}
Q.~Shi, L.~P. Chen, G.~J. Nan, R.~X. Xu, and Y.~J. Yan,
\newblock J. Chem. Phys. {\bf 130}, 164518 (2009).

\bibitem{Zhu115678Note}
{Note the indexing error in the eq 18 of Ref.\ \cite{Zhu115678}.
}

\bibitem{Yan885160}
Y.~J. Yan and S.~Mukamel,
\newblock J. Chem. Phys. {\bf 89}, 5160 (1988).

\bibitem{Yan91179}
Y.~J. Yan and S.~Mukamel,
\newblock J. Chem. Phys. {\bf 94}, 179 (1991).

\bibitem{Muk02327}
S.~Mukamel and A.~Tortschanoff,
\newblock Chem. Phys. Lett. {\bf 357}, 327  (2002).

\bibitem{Muk07221105}
S.~Mukamel, R.~Oszwa{\l}dowski, and L.~Yang,
\newblock J. Chem. Phys. {\bf 127}, 221105 (2007).

\bibitem{Zhu115678}
K.~B. Zhu, R.~X. Xu, H.~Y. Zhang, J.~Hu, and Y.~J. Yan,
\newblock J. Phys. Chem. B {\bf 115}, 5678 (2011).

\end{thebibliography}

\end{document}